\documentclass{article}
\usepackage{graphicx}
\usepackage{amsmath}

\input{tcilatex}

\begin{document}

\title{\textbf{Free} \textbf{massless particles and extended}\\
\textbf{space-time algebra}}
\author{W. F. Chagas-Filho \\
Departamento de Fisica, Universidade Federal de Sergipe\\
Aracaju, SE, Brazil}
\maketitle

\begin{abstract}
We study the space-time invariances of the relativistic particle action for
both the massive and massless cases. While the massive action has only the
invariances associated to the Poincar\'{e} algebra, we find that the
invariances of the massless action give rise to the conformal algebra in
four dimensions. For the free massless particle, a new invariance of the
action permits the construction of an extension of the conformal algebra.
The conclusion is that two distinct symmetry breaking mechanisms are
necessary to arrive at the Poincar\'{e} algebra.

\end{abstract}

\section{Introduction}

The relativistic particle continues to be one of the most interesting
dynamical systems to investigate if one wishes to understand fundamental
physics. This is because relativistic particle theory has many features that
have higher-dimensional analogues in relativistic string theory, while, at
the same time, being a prototype of general relativity. In the well-known
``einbein'' version [1], the action for the relativistic particle 
\begin{equation}
S=\frac{1}{2}\int d\tau (e^{-1}\dot{x}^{2}-em^{2})  \tag{1}
\end{equation}
defines a generally covariant one-dimensional field theory where the
particle mass $m$ plays the role of a cosmological constant [2]. Action (1)
is then the simplest theoretical laboratory where we can investigate the
origin of a non-vanishing cosmological constant [3].

The classical equation of motion for $x^{\mu }$ that follows from Hamilton's
principle applied to action (1) is 
\begin{equation}
\frac{d}{d\tau }(e^{-1}\dot{x}^{\mu })=0  \tag{2}
\end{equation}
and the particle will be free only when the condition $\frac{de}{d\tau }=0$
is satisfied. As the variable $e(\tau )$ is associated to the geometry of
the particle world-line, this condition is equivalent to a constant,
non-dynamical, geometry. Let us study the space-time invariances of action
(1). It is invariant under Poincar\'{e} transformations 
\begin{equation}
\delta x^{\mu }=a^{\mu }+\omega _{\nu }^{\mu }x^{\nu }  \tag{3a}
\end{equation}
\begin{equation}
\delta e=0  \tag{3b}
\end{equation}
and under the diffeomorphisms 
\begin{equation}
\delta x^{\mu }=\dot{\epsilon}x^{\mu }  \tag{4a}
\end{equation}
\begin{equation}
\delta e=\frac{d}{d\tau }(\epsilon e)  \tag{4b}
\end{equation}
In consequence of the invariance of action (1) under the Poincar\'{e}
transformation (3), the following vector field is defined on the background
space-time [4] 
\begin{equation}
V=a^{\mu }P_{\mu }-\frac{1}{2}\omega ^{\mu \nu }M_{\mu \nu }  \tag{5}
\end{equation}
where 
\begin{equation}
P_{\mu }=\partial _{\mu }  \tag{6}
\end{equation}
\begin{equation}
M_{\mu \nu }=x_{\mu }\partial _{\nu }-x_{\nu }\partial _{\mu }  \tag{7}
\end{equation}
$a^{\mu }P_{\mu }$ is the field of space-time translations and $\frac{1}{2}%
\omega ^{\mu \nu }M_{\mu \nu }$ is the field of space-time rotations. The
generators of these fields obey the algebra 
\begin{equation}
\lbrack P_{\mu },P_{\nu }]=0  \tag{8a}
\end{equation}
\begin{equation}
\lbrack P_{\mu },M_{\nu \lambda }]=\delta _{\mu \nu }P_{\lambda }-\delta
_{\mu \lambda }P_{\nu }  \tag{8b}
\end{equation}
\begin{equation}
\lbrack M_{\mu \nu },M_{\lambda \rho }]=\delta _{\nu \lambda }M_{\mu \rho
}+\delta _{\mu \rho }M_{\nu \lambda }-\delta _{\nu \rho }M_{\mu \lambda
}-\delta _{\mu \lambda }M_{\nu \rho }  \tag{8c}
\end{equation}
The algebra (8) is the Poincar\'{e} space-time algebra.

\section{Massless particles}

Let us now consider the massless particle action 
\begin{equation}
S=\frac{1}{2}\int d\tau e^{-1}\dot{x}^{2}  \tag{9}
\end{equation}
which is the $m=0$ limit of action (1). This action is also invariant under
the Poincar\'{e} transformations (3) and under the diffeomorphisms (4). The
equation of motion for $x^{\mu }$ that follows from (9) is identical to (2),
and the massless particle will only be free if the one-dimensional geometry
defined by the particle world-line is a non-dynamical one.

The massless action (9) has a larger set of space-time invariances. It is
also invariant under the scale transformation 
\begin{equation}
\delta x^{\mu }=\alpha x^{\mu }  \tag{10a}
\end{equation}
\begin{equation}
\delta e=2\alpha e  \tag{10b}
\end{equation}
where $\alpha $ is a constant, and under the conformal transformation 
\begin{equation}
\delta x^{\mu }=(2x^{\mu }x^{\nu }-\eta ^{\mu \nu }x^{2})b_{\nu }  \tag{11a}
\end{equation}
\begin{equation}
\delta e=4ex.b  \tag{12b}
\end{equation}
where $b_{\mu }$ is a constant vector. As a consequence of the invariances
of the massless action we can define the space-time vector field [4] 
\begin{equation}
V_{0}=a^{\mu }P_{\mu }-\frac{1}{2}\omega ^{\mu \nu }M_{\mu \nu }+\alpha
D+b^{\mu }K_{\mu }  \tag{13}
\end{equation}
where 
\begin{equation}
D=x^{\mu }\partial _{\mu }  \tag{14}
\end{equation}
and 
\begin{equation}
K_{\mu }=(2x_{\mu }x^{\nu }-\delta _{\mu }^{\nu }x^{2})\partial _{\nu } 
\tag{15}
\end{equation}
The two additional vector fields on the right of equation (13) are
respectively associated with dilatations and conformal boosts. The
generators of the vector field $V_{0\text{ \ }}$obey the algebra 
\begin{equation}
\lbrack P_{\mu },P_{\nu }]=0  \tag{16a}
\end{equation}
\begin{equation}
\lbrack P_{\mu },M_{\nu \lambda }]=\delta _{\mu \nu }P_{\lambda }-\delta
_{\mu \lambda }P_{\nu }  \tag{16b}
\end{equation}
\begin{equation}
\lbrack M_{\mu \nu },M_{\lambda \rho }]=\delta _{\nu \lambda }M_{\mu \rho
}+\delta _{\mu \rho }M_{\nu \lambda }-\delta _{\nu \rho }M_{\mu \lambda
}-\delta _{\mu \lambda }M_{\nu \rho }  \tag{16c}
\end{equation}
\begin{equation}
\lbrack D,D]=0  \tag{16d}
\end{equation}
\begin{equation}
\lbrack D,P_{\mu }]=-P_{\mu }  \tag{16e}
\end{equation}
\begin{equation}
\lbrack D,M_{\mu \nu }]=0  \tag{16f}
\end{equation}
\begin{equation}
\lbrack D,K_{\mu }]=K_{\mu }  \tag{16g}
\end{equation}
\begin{equation}
\lbrack P_{\mu },K_{\nu }]=2(\delta _{\mu \nu }D-M_{\mu \nu })  \tag{16h}
\end{equation}
\begin{equation}
\lbrack M_{\mu \nu },K_{\lambda }]=\delta _{\nu \lambda }K_{\mu }-\delta
_{\lambda \mu }K_{\nu }  \tag{16i}
\end{equation}
\begin{equation}
\lbrack K_{\mu },K_{\nu }]=0  \tag{16j}
\end{equation}
This is the conformal algebra in a $D=4$ space-time [4], the extension of
the Poincar\'{e} algebra (8).

Let us now restrict the analysis to the case of a non-dynamical
one-dimensional geometry. Using the equation of free motion, it can be
verified that the massless particle action (9) is also invariant under the
transformation [5] 
\begin{equation}
x^{\mu }\rightarrow \exp \{\frac{1}{3}\beta (\dot{x}^{2})\}x^{\mu } 
\tag{17a}
\end{equation}
\begin{equation}
e\rightarrow \exp \{\frac{2}{3}\beta (\dot{x}^{2})\}e  \tag{17b}
\end{equation}
where $\beta $ is an arbitrary function of $\dot{x}^{2}$. This symmetry is
interesting because it has a higher-dimensional extension in the tensionless
limit of string theory [5]. Just as the massless limit is the high-energy
limit of particle theory, the tensionless limit [6] is the high-energy limit
[7] of string theory. Here, if $\tau $ is taken to be the particle's proper
time, then $\dot{x}^{\mu }$ is the four-velocity and equations (17) define a
scale transformation that depends on the square of the four-velocity.
Infinitesimally we can then define velocity-dependent scale transformations 
\begin{equation}
\delta x^{\mu }=\alpha \beta (\dot{x}^{2})x^{\mu }  \tag{18}
\end{equation}
where $\alpha $ is the same constant that appears in equations (10). These
transformations then lead to the existence of velocity-dependent
dilatations. The vector field $D$ of equation (14) can then be changed
according to 
\begin{equation}
D=x^{\mu }\partial _{\mu }\rightarrow D^{\ast }=x^{\mu }\partial _{\mu
}+\beta (\dot{x}^{2})x^{\mu }\partial _{\mu }  \tag{19}
\end{equation}
Because all vector fields in equation (13) involve partial derivatives with
respect to $x^{\mu }$ and $\beta $ is a function of $\dot{x}^{\mu }$, we can
also introduce the generators 
\begin{equation}
P_{\mu }^{\ast }=P_{\mu }+\beta P_{\mu }  \tag{20}
\end{equation}
\begin{equation}
M_{\mu \nu }^{\ast }=M_{\mu \nu }+\beta M_{\mu \nu }  \tag{21}
\end{equation}
\begin{equation}
K_{\mu }^{\ast }=K_{\mu }+\beta K_{\mu }  \tag{22}
\end{equation}
and define a new vector field $V_{0}^{\ast }$ \ by 
\begin{equation}
V_{0}^{\ast }=a^{\mu }P_{\mu }^{\ast }-\frac{1}{2}\omega ^{\mu \nu }M_{\mu
\nu }^{\ast }+\alpha D^{\ast }+b^{\mu }K_{\mu }^{\ast }  \tag{23}
\end{equation}
The generators of this vector field obey the algebra 
\begin{equation}
\lbrack P_{\mu }^{\ast },P_{\nu }^{\ast }]=0  \tag{24a}
\end{equation}
\begin{equation}
\lbrack P_{\mu }^{\ast },M_{\nu \lambda }^{\ast }]=(\delta _{\mu \nu
}P_{\lambda }^{\ast }-\delta _{\mu \lambda }P_{\nu }^{\ast })+\beta (\delta
_{\mu \nu }P_{\lambda }^{\ast }-\delta _{\mu \lambda }P_{\nu }^{\ast }) 
\tag{24b}
\end{equation}
\begin{eqnarray}
\lbrack M_{\mu \nu }^{\ast },M_{\lambda \rho }^{\ast }] &=&(\delta _{\nu
\lambda }M_{\mu \rho }^{\ast }+\delta _{\mu \rho }M_{\nu \lambda }^{\ast
}-\delta _{\nu \rho }M_{\mu \lambda }^{\ast }-\delta _{\mu \lambda }M_{\nu
\rho }^{\ast })  \notag \\
&&+\beta (\delta _{\nu \lambda }M_{\mu \rho }^{\ast }+\delta _{\mu \rho
}M_{\nu \lambda }^{\ast }-\delta _{\nu \rho }M_{\mu \lambda }^{\ast }-\delta
_{\mu \lambda }M_{\nu \rho }^{\ast })  \TCItag{24c}
\end{eqnarray}
\begin{equation}
\lbrack D^{\ast },D^{\ast }]=0  \tag{24d}
\end{equation}
\begin{equation}
\lbrack D^{\ast },P_{\mu }^{\ast }]=-P_{\mu }^{\ast }-\beta P_{\mu }^{\ast }
\tag{24e}
\end{equation}
\begin{equation}
\lbrack D^{\ast },M_{\mu \nu }^{\ast }]=0  \tag{24f}
\end{equation}
\begin{equation}
\lbrack D^{\ast },K_{\mu }^{\ast }]=K_{\mu }^{\ast }+\beta K_{\mu }^{\ast } 
\tag{24g}
\end{equation}
\begin{equation}
\lbrack P_{\mu }^{\ast },K_{\nu }^{\ast }]=2(\delta _{\mu \nu }D^{\ast
}-M_{\mu \nu }^{\ast })+2\beta (\delta _{\mu \nu }D^{\ast }-M_{\mu \nu
}^{\ast })  \tag{24h}
\end{equation}
\begin{equation}
\lbrack M_{\mu \nu }^{\ast },K_{\lambda }^{\ast }]=(\delta _{\lambda \nu
}K_{\mu }^{\ast }-\delta _{\lambda \mu }K_{\nu }^{\ast })+\beta (\delta
_{\lambda \nu }K_{\mu }^{\ast }-\delta _{\lambda \mu }K_{\nu }^{\ast }) 
\tag{24i}
\end{equation}
\begin{equation}
\lbrack K_{\mu }^{\ast },K_{\nu }^{\ast }]=0  \tag{24j}
\end{equation}
Notice that the vanishing brackets of the conformal algebra (16) are
preserved as vanishing in the above algebra, but the non-vanishing brackets
of the conformal algebra now have linear and quadratic contributions from
the arbitrary function $\beta (\dot{x}^{2})$. It is interesting to choose $%
\beta $ simply linear in $\dot{x}^{2}$ because then, if the classical
equation of motion for $e(\tau )$ that follows from the massless action (9)
is imposed, the transformation (17) becomes the identity transformation. The
algebra (24) is then on-shell in $x^{\mu }$ but off-shell in $e$.

The conclusions of this letter are the following: For the relativistic
particle, two different kinds of symmetry breaking mechanisms are necessary
to go from the algebra (24) to the Poincar\'{e} algebra (8). The first one
is the appearance of an interaction. When this occurs, the free motion
equation is no longer valid and the invariance (17) disappears. The algebra
(24) then reduces to the conformal algebra (16). The second mechanism must
generate a non-vanishing particle mass. This will destroy scale invariance
and conformal invariance, and the conformal algebra (16) finally reduces to
the Poincar\'{e} algebra (8). These features of relativistic particle theory
may have some significance in relation to the cosmological constant problem
because, as we mentioned in the introduction, the particle mass plays the
role of a cosmological constant in the one-dimensional generally covariant
field theory defined by action (1). According to the results of this letter,
a non-vanishing cosmological constant may be the result of symmetry breaking
mechanisms.

\end{document}